# Study of the Physical Properties of the $EuCo_2As_2$ Compound: A DFT approach

S. Benyoussef [1], A. Jabar [2,3] and L. Bahmad [1,*]

[1] Laboratory of Condensed Matter and Interdisciplinary Sciences (LaMCScI), Faculty of Sciences, Mohammed V University, Av. Ibn Batouta, B. P. 1014 Rabat, Morocco

[2] LPMAT, Faculty of Sciences Aïn Chock, Hassan II University of Casablanca, B.P. 5366 Casablanca, Morocco.

[3] LPHE-MS, Science Faculty, Mohammed V University in Rabat, Morocco

*Corresponding author: l.bahmad@um5r.ac.ma (L.B.)

**Abstract**

In this study, we carried out an investigation of the $EuCo_2As_2$ compound, focusing on its various physical properties. Our analysis covered the structural, magnetic, electronic, optical, thermodynamic and thermoelectric characteristics of this compound. To carry out this study, we used density functional theory (DFT) implemented in the Wien2k software package. To determine the exchange-correlation potential, we used the GGA-PBE (Perdew, Burke and Ernzerhof) approach, taking spin-orbit coupling (SOC) into account. Our results indicate that the $EuCo_2As_2$ compound exhibits metallic behavior. In addition, we have determined that the compound's stable ground state is the ferromagnetic (FM) phase. We have also calculated the Debye temperature and the Grüneisen parameter. In addition, we evaluated various optical properties, including electron energy loss, absorption coefficient, real and imaginary dielectric tensors, and real and imaginary optical conductivity. We found that the compound has excellent absorption characteristics in the low and mid ultraviolet (UV) spectra. In addition, we investigated the electrical conductivity, Seebeck coefficient, electronic conductivity and thermal conductivity of the lattice. The results revealed that the compound exhibits n-type behavior, with negative values for the Seebeck coefficient. These results are analyzed in detail and provide valuable information on the properties of the $EuCo_2As_2$ compound. Additionally, the computed parameters were compared to those found in the literature. A good deals have been revealed with the existing results.

## 1. Introduction

In recent years, there has been a growing interest in utilizing pressure as a variable to investigate and discover various properties in materials, such as superconductivity, magnetic behavior, and structural phase transitions. $EuCo_2As_2$ is a member of a group of materials that have received significant attention named the $RT_2X_2$ pnictides, where R represents rare-earth or alkaline-earth metals, T denotes transition metals, and X represents pnictogens. These pnictides crystallize in the tetragonal $ThCr_2Si_2$-type structure, known for its diverse range of intriguing physical phenomena, including unconventional superconductivity, heavy fermion states, valence fluctuations, quantum criticality, and multiple magnetic transitions [1, 2]. Among these pnictides, the $AFe_2As_2$ compounds, with A representing divalent alkaline-earth or rare-earth metals, exhibit superconductivity under specific conditions, such as chemical doping or application of external pressure, while adopting the tetragonal $ThCr_2Si_2$ ("122")-type structure [3]. In ref. [4], theoretical calculations indicate that $EuCo_2As_2$ undergoes a structural phase transition from the tetragonal phase to the collapsed tetragonal phase under the pressure of approximately 4 GPa. Researchers have studied the crystal structure and electrical resistance of structurally layered $EuFe_2As_2$ up to 70 GPa and down to a temperature of 10 K. The results demonstrate that the tetragonal phase of $EuFe_2As_2$ (I4/mmm) leads to an increase in the length of the a-axis and a significant decrease in the length of the c-axis as pressure increases [5]. In a separate study [6], fifteen new compounds of the $AB_2X_2$ type were synthesized, where A represents a lanthanoid, B denotes Fe, Co, or Ni, and X represents P, As, or Sb. These compounds crystallize with the $ThCr_2Si_2$-type structure. This reveals a phase transition from a tetragonal phase to a collapsed tetragonal phase in $EuCo_2As_2$ at 4.7 GPa [7]. In a previous study [3], the well-known $EuFe_2As_2$ ferro-pnictide compound exhibited intriguing spin and charge dynamics, resulting in the coexistence of superconductivity and magnetism under specific conditions. More recently, it has been discovered that applying pressure to the initially antiferromagnetic compound $EuCo_2As_2$ ($T_N$ = 47 K) induces ferromagnetism [9]. This change in magnetic ordering is associated with the stabilization of a mixed valence state of europium. Another study [10] demonstrated a significant modification of antiferromagnetism in $PrCo_2P_2$ through alio-valent substitution within the Pr sub-lattice. Both $Pr_{0.8}Eu_{0.2}Co_2P_2$ and $Pr_{0.8}Ca_{0.2}Co_2P_2$ undergo a ferromagnetic phase transition due to the ordering of cobalt magnetic moments near 280 K. Furthermore, $EuCo_2P_2$ and $SrNi_2P_2$ exhibit a first-order phase transition, characterized by substantial hysteresis and highly pronounced, anisotropic changes in the lattice parameters at pressures of 30 kBar and 4 kBar, respectively [11].

In contrast to compounds like $EuFe_2P_2$, which undergo a second-order phase transition under pressure as shown in references [11, 12], compounds such as $EuCo_2P_2$, where cobalt replaces iron, exhibit a first-order pressure-induced structural phase transition. A study on $EuFe_2A_2$ in ref. [12] reveals a phase transition accompanied by an extremely anisotropic and negative compressibility axial phenomenon, where the length of the a-axis increases while the length of the c-axis decreases under pressure in ternary iron pnictide superconductors. Single-crystal magnetization measurements conducted on $EuCo_2As_2$ at ambient pressure showed that the effective paramagnetic moment at high temperatures primarily arises from the $Eu^{2+}$ spins, while below the ordering temperature, a strong anisotropic behavior was observed in both magnetic field and temperature dependencies [1].

The anisotropic magnetic properties of the layered helimagnet $EuCo_2As_2$ have been investigated, providing insights into the evolving spin configurations under rotating magnetic fields [13]. In the rare-earth compound $EuCu_2Ge_2$, it has been observed that the europium valence reaches 2.89 at pressures exceeding 20 GPa [2]. Despite the presence of significant ferromagnetic low-energy spin excitations, $BaCo_2As_2$ and $SrCo_2As_2$ remain in a paramagnetic state even at very low temperatures. This behavior can be attributed to the weaker electronic correlation strength of the Co 3d $x^2$-$y^2$ orbital and its distance from the Van Hove singularity [14]. Single crystalline samples of $Ba(Fe_{1-x}Co_x)_2As_2$ with x<0.12 exhibit magnetic transitions from a high-temperature nonmagnetic phase to a low-temperature antiferromagnetic state [15]. In the case of As-based Zintl compounds $Ba_{1-x}K_xCd_2As_2$, which adopt the $CaAl_2Si_2$-type structure (space group $P\bar{3}m1$), both the electrical resistivity and the Seebeck coefficient decrease with hole doping. This trend leads to a power factor value of 1.28 $W.m^{-1}.K^{-2}$ at 762 K for x = 0.04. Consequently, the dimensionless figure-of-merit ZT reaches its maximum value of 0.81 at 762 K for x = 0.04 [16]. Recent theoretical investigations on various materials have been published using ab-initio to determine the electronic and optical properties while the semi-classical Boltzmann transport theory within the BoltzTraP code was used to demonstrate the thermoelectric properties [17-21] where the findings match with the experimental results.

To the best of our knowledge, no theoretical research has been conducted on all aspects of the physical properties of the $EuCo_2As_2$ compound. Therefore, in this work, physical properties such as structural, electronic, optical, thermodynamic and thermoelectric properties will be studied to fill the gaps in the literature. Therefore, this paper is organized as follows: in Section 2, we describe the details of the computational method. In Section 3, we examine the obtained results for the structural, electronic, magnetic, optical, thermodynamic, and thermoelectric properties of the studied materials. Section 4 is devoted to conclusions.

## 2. DFT methodology

The electronic, optical, and magnetic properties of the system were investigated using the WIEN2K program, which is based on the full-potential linearized augmented plane wave (FPLAPW) method with a dual basis set [22]. The exchange-correlation potential in the FPLAPW calculations was treated using the GGA-PBE (Perdew, Burke, and Ernzerhof) functional [22]. The size of the basis sets was determined by the parameter $R_{MT} \times K_{max}$, where $R_{MT}$ represents the smallest muffin tin radius in the unit cell and $K_{max}$ is the magnitude of the largest K vector in reciprocal space. In our calculations, the basic functions were expanded up to $R_{MT} \times K_{max} = 9$, and a total of 1000 points were used for Brillouin zone integration, employing a modified tetrahedron method. To account for electron-electron Coulomb interactions in the Eu and Co atoms, the GGA+U approach was employed, considering the rotationally invariant ways [22]. In this study, the effective parameter $U_{eff} = U - J$ was adopted, where U and J represent the Coulomb and exchange parameters, respectively. For simplicity, we use U to denote the effective parameter in the subsequent discussion. The self-consistent calculations were considered convergent when the energy convergence was below $10^{-5}$ Ry. The inclusion of spin-orbit coupling (SOC) was carried out using the second-variational method with scalar relativistic wave functions [22]. For the calculation of thermoelectric coefficients, a combination of first-principles band structure calculations and the Boltzmann transport theory within the rigid band approximation (RBA) and the constant scattering time approximation (CSTA) was employed, utilizing the BoltzTrap code [23].

$EuCo_2As_2$ belongs to the $ThCr_2Si_2$-type structure family, adopting a body-centered tetragonal structure with lattice constants of a = 3.9752 Å and c = 11.1011 Å, as illustrated in Fig. 1. The crystalline arrangement in the $ThCr_2Si_2$-type structure consists of Eu atoms at the 2a position (0,0,0), Co atoms at the 4d positions (0,1/2,1/4) and (1/2, 0, 1/4), and As atoms at the 4e positions (0,0,z) and (0,0,-z) [3, 4].

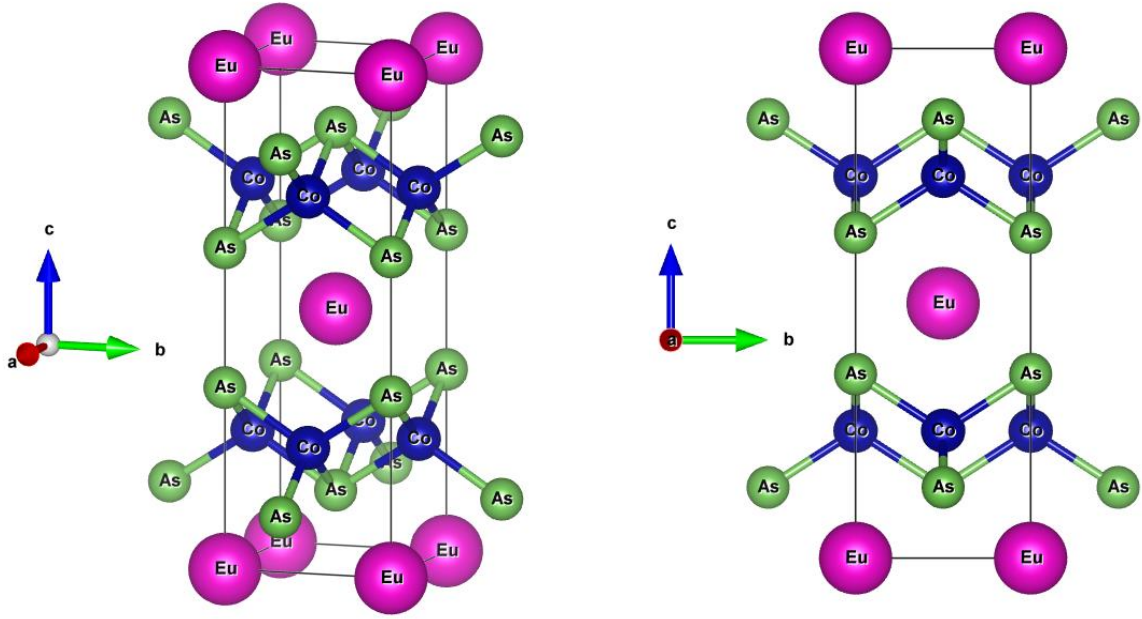

*Fig. 1*: *The crystal structure of EuCo2As2*

### 3. Results and discussions

#### 3.1. Total energy and lattice parameters

Before investigating the electronic transport properties of $EuCo_2As_2$, a structural optimization is conducted to determine the lattice constants, bulk modulus, and static pressure transitions for both the Ferromagnetic (FM) and Antiferromagnetic (AFM) orderings. The optimization is performed using the empirical Birch-Murnaghan equation of state [24].

The optimized parameters for the studied compound in the FM and AFM states are obtained, including the lattice constants ($a_0$ and $c_0$), volume ($V_0$), bulk modulus ($B_0$), its first pressure derivative (B'), and minimum total energy ($E_0$). The values of these parameters are summarized in Table 1. Additionally, the energy as a function of the volume curves is depicted in Fig.2. The results clearly demonstrate that the energy ordering of these phases, as predicted by the FP-LAPW method, follows the pattern $E_{FM} < E_{AFM}$. This indicates that the FM phase is energetically more favorable than the AFM phase, establishing it as the stable ground state of the $EuCo_2As_2$ compound. From this table, a good agreement was observed between the present findings and experiment results [13] where the ferromagnetic order has been found.

**Table 1:** Calculated equilibrium lattice parameters, such as lattice constants $a_0$ and $c_0$ (in Å), volume $V_0$ (in Å$^3$), bulk modulus $B_0$ (in GPa), its first pressure derivative B', and the minimum total energy $E_0$ (in Ry) for the tetragonal EuCo$_2$As$_2$ compound, antiferromagnetic (AFM), and ferromagnetic (FM) states, by employing the GGA-PBE approximation.

|     | $a_0$(Å) | $c_0$(Å) | $V_0$(Å$^3$) | B(GPa) | B' | $E_0$(Ry) |
|-----|----------|----------|--------------|--------|-----|-----------|
| FM  | 4.03     | 10.907   | 177.711      | 107.615| 4.811 | -36323.53486 |
| AFM | 4.03     | 10.907   | 177.711      | 108.302| 4.805 | -36323.53270 |

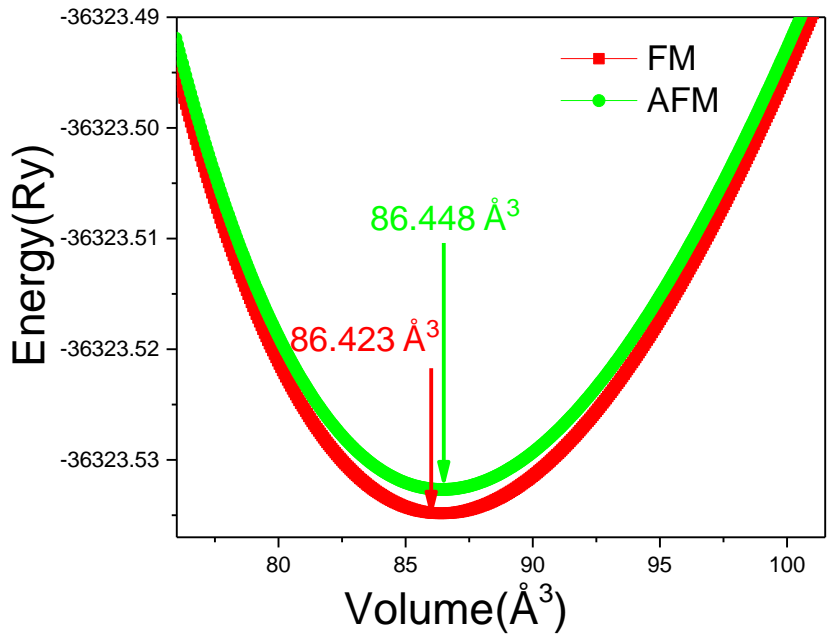

***Fig. 2***: *The energy vs. the volume of the EuAs2Co2 compound.*

### 3.2. Elastic constants

The elastic stiffness constants play a crucial role as they are connected to inter-atomic bonding, phonon spectra, and various fundamental solid-state phenomena. Furthermore, they are closely linked to mechanical properties and micro-hardness. In many solids, the stress-strain curve exhibits minimal nonlinearity, allowing for the assumption of a linear relationship between stress and strain in practical applications.

To predict the elastic constants of tetragonal EuCo$_2$As$_2$ alloys, a numerical first-principles calculation method developed by Thomas Charpin was utilized, which is integrated into the Wien2k package, as described in ref. [25]. This method involves applying small strains to the unstrained lattice to calculate the elastic constants. For a tetragonal symmetry, there are six independent elastic constants denoted as $C_{11}$, $C_{12}$, $C_{13}$, $C_{33}$, $C_{44}$, and $C_{66}$. To determine these constants, six different strains are required. The calculated values of the $C_{ij}$ constants are provided in Table 2. These computed elastic moduli values for the tetragonal $EuCo_2As_2$ compound serve as a fundamental reference for future projects involving $EuCo_2As_2$ material.

**Table 2:** The calculated elastic constants $C_{ij}$ in the unit of (GPa) for $EuCo_2As_2$.

| $C_{11}$ | $C_{12}$ | $C_{13}$ | $C_{33}$ | $C_{44}$ | $C_{66}$ |
|---|---|---|---|---|---|
| 241.0 | -17.3 | 79.2 | 79.9 | 126.3 | 24.6 |

### 3.3. Magnetic properties

In this part, the magnetic properties of $EuCo_2As_2$ have been investigated [26]. The transition temperature ($T_C$) is determined by employing the mean-field approach (MFA). This parameter could be computed by the relation [27]:

$$T_C = \frac{2 \times \Delta E}{3 k_B} \quad (1)$$

Where $\Delta E$ and $k_B$ refer to the difference in total energy between the ferromagnetic and antiferromagnetic states and the Boltzmann constant, respectively. The calculated Curie temperature values, and the total and the partial magnetic moments of $EuCo_2As_2$ are listed in Table 3.

The electron-electron Coulomb interaction (U) plays a crucial role in understanding the properties of transition-metal oxides, especially the 3d transition-metal oxides. However, determining an appropriate value for U that accurately reproduces experimental data is challenging. This is due to the fact that the parameter U is system-specific and its dependence on the surrounding environment is not well understood. In general, applying the U parameter helps distinguish between occupied and unoccupied states of the atom to which it is applied. Therefore, in order to investigate the effect of spin-orbit coupling (SOC) and Hubbard potential (GGA+SOC+U) on the magnetic and electronic properties of $EuCo_2As_2$, $U_{eff}$ = U-J = 4.5 eV was performed. It is seen that GGA-PBE gives the magnetic moment of Eu at 6.72112 $\mu_B$, exactly the same as the value obtained from the GGA+SOC+U calculation, which is 6.75388 $\mu_B$. The magnetic moment on Co was observed with the GGA-PBE calculation of 0.82555 $\mu_B$ while our GGA+SOC+U gives a larger value of 2.44753 $\mu_B$. For the As atom, GGA+SOC+U gives an orbital moment of 0.04044 $\mu_B$, much larger than the -0.01960 $\mu_B$ of the GGA-PBE calculation. The value of Curie temperature is found to be 227.253 K, a result which is in agreement with the Ref. [13].

**Table 3:** The total and partial magnetic moments (in $\mu_B$) and the Curie temperature (in K) of the $EuCo_2As_2$ compound.

|  | $M_{Eu}(\mu_B)$ | $M_{Co}(\mu_B)$ | $M_{As}(\mu_B)$ | $T_C(K)$ |
|---|---|---|---|---|
| GGA-PBE | 6.72112 | 0.82555 | -0.01960 | 227.253 |
| GGA+SOC+U | 6.75388 | 2.44753 | 0.04044 |  |

### 3.4. Electronic properties

Since the $EuCo_2As_2$ is a new material, our research first focuses on a comprehensive analysis of the electronic structure and bonding using density functional theory (DFT) calculations. The band structure of $EuCo_2As_2$ along the high-symmetry lines in the Brillouin Zone as determined by GGA-PBE and GGA+SOC+U is shown in Fig. 3 for both spins (up and down). The computation was performed in the first Brillouin zone on the path Γ–H–N–Γ–P. In these plots, bands are identified for the investigation to be done on this metallic compound. So, this crossing of the Fermi level by bands is directly showing the existence of the metallic nature of this compound, which is in accordance with the experimentally observed metallic state [7].

One of the most important variables in determining the electronic characteristics of this material is the electronic density of states (DOS). The total density of states (TDOS) and partial density of states (PDOS) of Eu, Co, and As have been calculated, as shown in Fig. 4 using GGA-PBE and GGA+SOC+U approaches. These diagrams show the participation of different atoms in the band structure as well as any possible linear combinations of their atomic orbitals. For both approaches, Eu and Co states predominate in the valence and conduction bands, while As states make a sizeable contribution in both the valence and conduction bands. The metallic character of the compound is demonstrated by the lack of a band gap, which is consistent with the conclusion drawn from the band structure shown in Fig.4. The asymmetric spin-up and spin-down contributions in TDOS also make the compound magnetic. These findings are in good agreement with the band structure's first-principle computations and density functional theory values reported in the literature, as well as being in excellent accord with earlier experimental Refs. [3, 9].

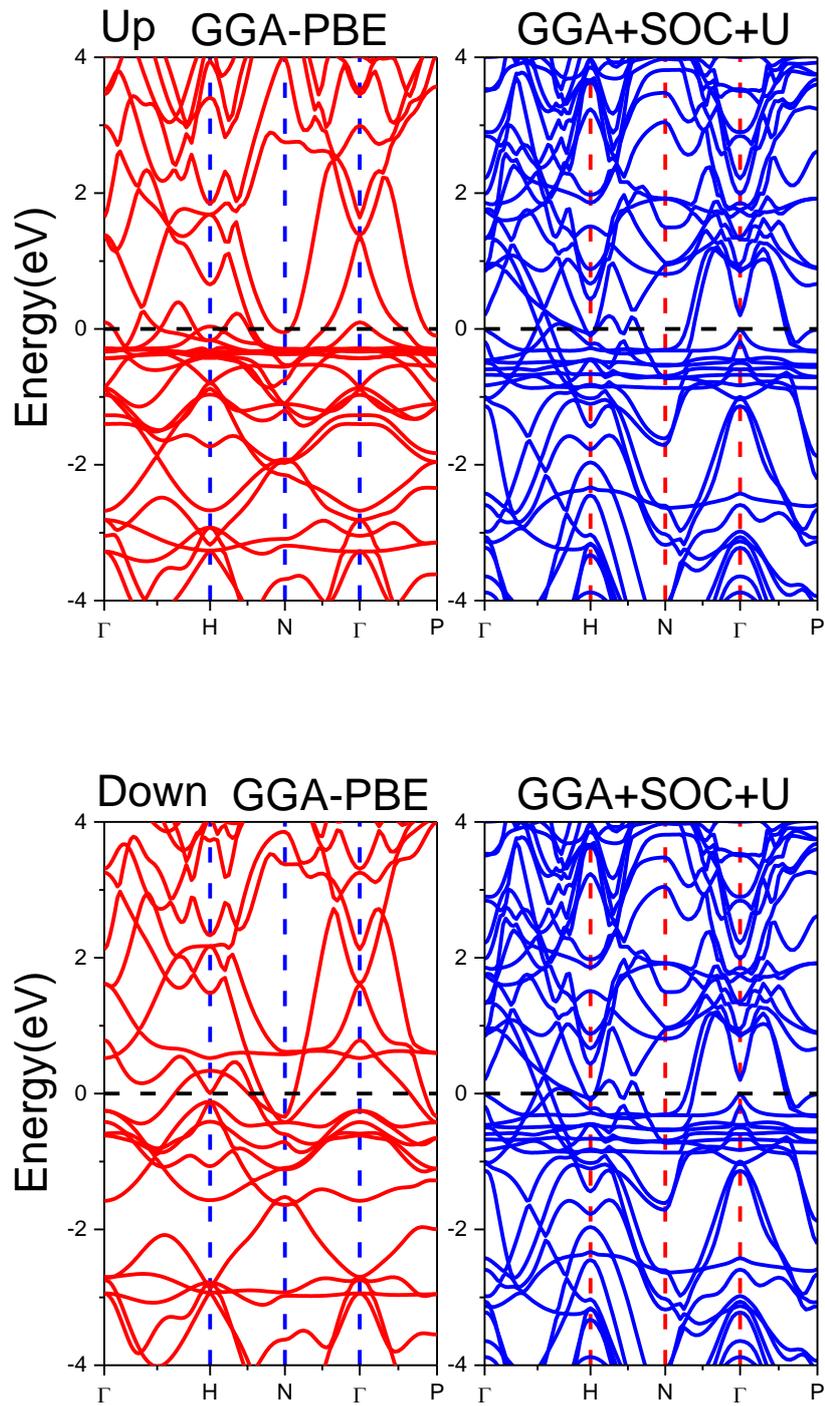

***Fig.3***: *The band structure up and down of EuAs$_2$Co$_2$ using GGA-PBE and GGA+SOC+U approaches.*

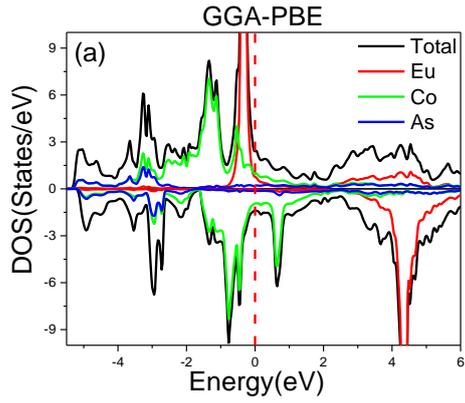
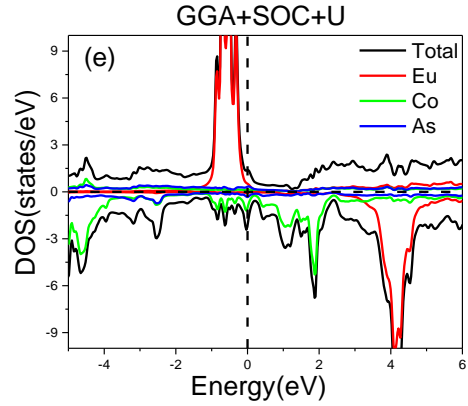
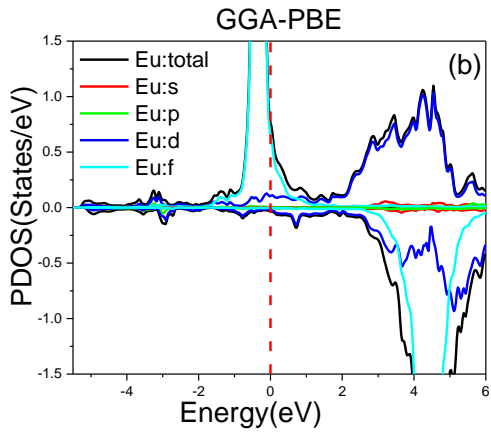
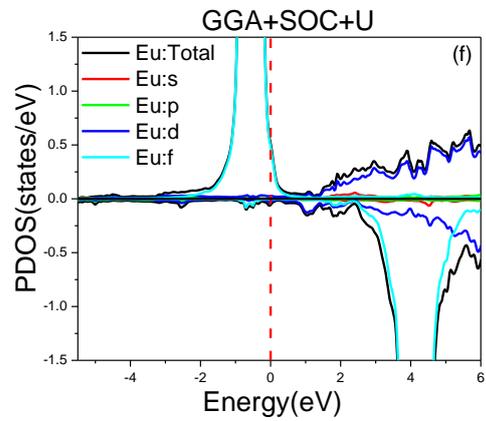
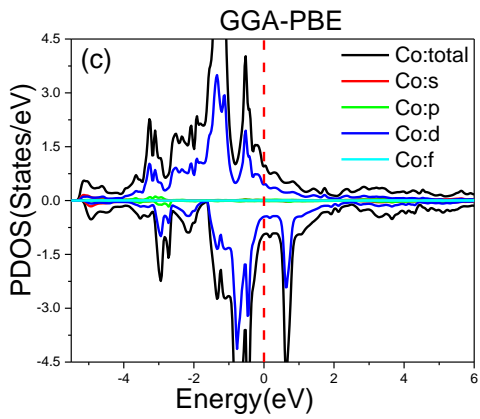
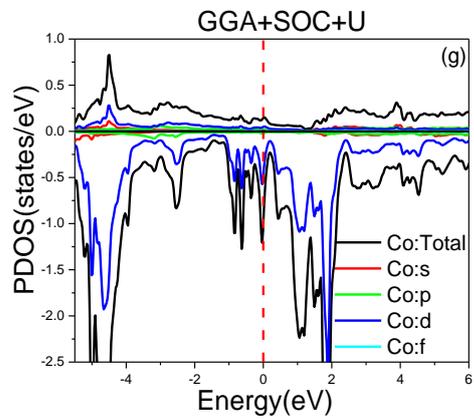

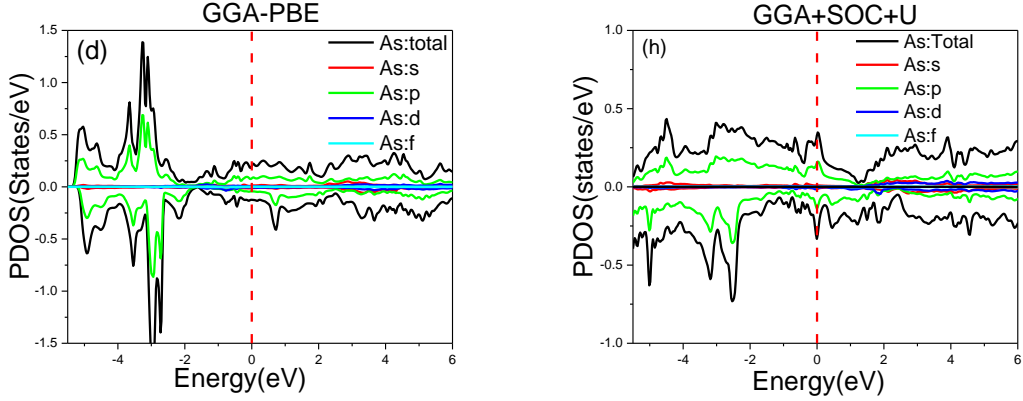

***Fig.4:*** *The total density of state (DOS) (a and e) and partial density of state (PDOS) (b, c, d, f, g, and h) of EuAs$_2$Co$_2$ using GGA-PBE and GGA+SOC+U approaches.*

### 3.5. Optical properties

The investigation of the optical properties of EuAs$_2$Co$_2$ is of great significance for exploring their potential applications in photovoltaic devices. To analyze the optical characteristics, various parameters such as energy electron loss, refractive index, extinction coefficient, absorption coefficient, and the real and imaginary parts of the dielectric tensor and optical conductivity are examined. Figs. 5 presents the results obtained from the study of these optical properties using both the GGA-PBE and GGA+SOC+U approaches, as discussed in references [28, 29]. The peak in the energy loss function reflects the dual nature of plasma resonance, which includes the associated frequency called plasma frequency. Above the plasma frequency, the material behaves as a dielectric, while below it exhibits metallic characteristics. This peak provides valuable information about the transition between the dielectric and metallic behavior of the material. The electron energy loss function for EuAs$_2$Co$_2$ is presented in Fig. 5(a) in the xx, yy, and zz directions. The highest energy loss function occurs in the ultraviolet region. It is well known that these peaks are due to inter-band transitions between various high symmetry points. The curve of EuAs$_2$Co$_2$ using GGA-PBE along zz direction is found to give the highest peak at an energy of 0.49 eV and the corresponding energy loss function value is 11 eV. Meanwhile, it is worth mentioning that this compound could be a good absorber of the low and medium UV spectrum. The refractive index is a dimensionless quantity that quantifies the transparency of a material to incident photons. It represents the speed at which light propagates through the material, providing a measure of how much the light is bent or refracted when passing through the material. Fig. 5(b) shows the refractive index for the xx, yy, and zz directions of EuAs$_2$Co$_2$. All curves decrease by reason of the increase in energy. But, when the energy of the incident photon reaches the value of 3.3 eV, the refractive index increases and reaches the maximum value of 4.4 for the xx, yy direction using the GGA+SOC+U approach.

The optical energy that is absorbed in the optical medium during light propagation is described by the extinction coefficient. As illustrated in Fig. 5(c), of extinction coefficients are presented for the xx, yy, and zz directions of EuAs$_2$Co$_2$. All curves decrease with increasing energy. We can see that the directions xx, yy, and zz give the same shape of the extinction coefficient. The absorption coefficient is a crucial optical property that plays a significant role in energetic materials and solar cells. It provides valuable insights into the extent to which light of a specific wavelength, with a well-defined energy can penetrate a material before being absorbed. This property is particularly relevant for optimizing the efficiency of solar energy conversion. As the sun emits light at various frequencies in the form of photons, understanding the absorption coefficient helps us evaluate how effectively a material can harness solar energy. The behavior of the calculated absorption coefficient of EuAs$_2$Co$_2$ has been depicted in Fig. 5(d). It has been observed that the curves increase in intensity when photon energies are higher and show a single peak in the UV domain. The higher one observed is the xx, yy direction using the GGA+SOC+U approach, and its intensity is 225x10$^4$ cm$^{-1}$ at energy 7 eV. Particularly, high absorption and quick photo-response make this compound an excellent choice for photovoltaic applications.

High dielectric constants are essential for achieving high-efficiency solar cells because they directly influence the strength of the Coulomb interaction between electron-hole pairs, charge carriers, and fixed ionic charges within the lattice. The dielectric constant determines the magnitude of this interaction, playing a crucial role in various processes such as charge separation, transport, and recombination. By having a high dielectric constant, solar cells can effectively manipulate and control these interactions, leading to improved device performance and enhanced energy conversion efficiency. Figs. 5(e, f) show the real and imaginary parts of the dielectric function for the xx, yy, and zz directions of EuAs$_2$Co$_2$. The curves of the real dielectric function begin to rise until reach peaks and decrease to a negative value with increasing energy. From the imaginary part as shown in Fig.5(f), we can see the light absorption of curves is mainly located in the regions of UV (~ 4eV).

The property of the material known as optical conductivity determines the link between the amplitude of the inducing electric field and the current density induced at any given frequency. The real and imaginary optical conductivity parts for the xx, yy, and zz directions of EuAs$_2$Co$_2$ are shown in Figs. 5(g, h). In the real part, all the peaks show an increase till a maximum peak where the curve of the xx, yy direction using the GGA+SOC+U approach at $15.5 \times 10^{15} s^{-1}$ is the highest one. Following that, the lines begin to flatten out as the energy level rises. All the curves of the imaginary part of optical conductivity show negative values at low energy and increase to positive values. The graph demonstrates that light absorption causes an increase in conductivity in the UV limit. The characteristic could be used in material photocells.

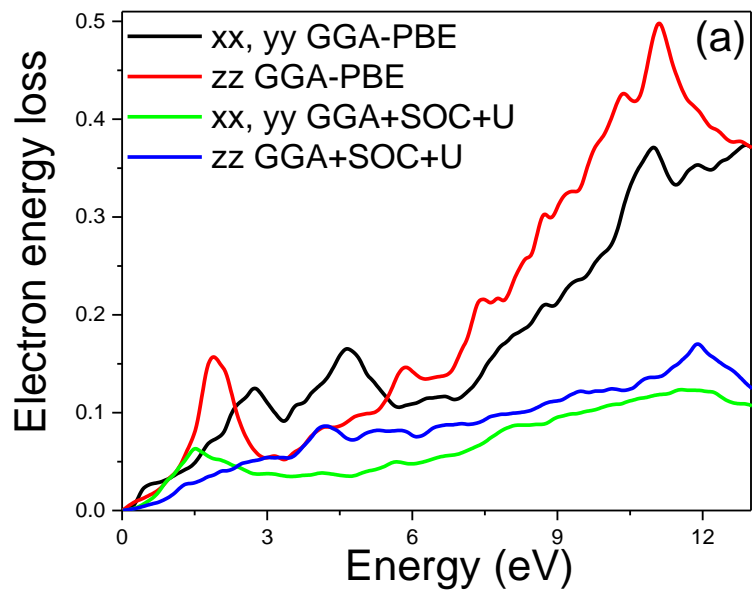

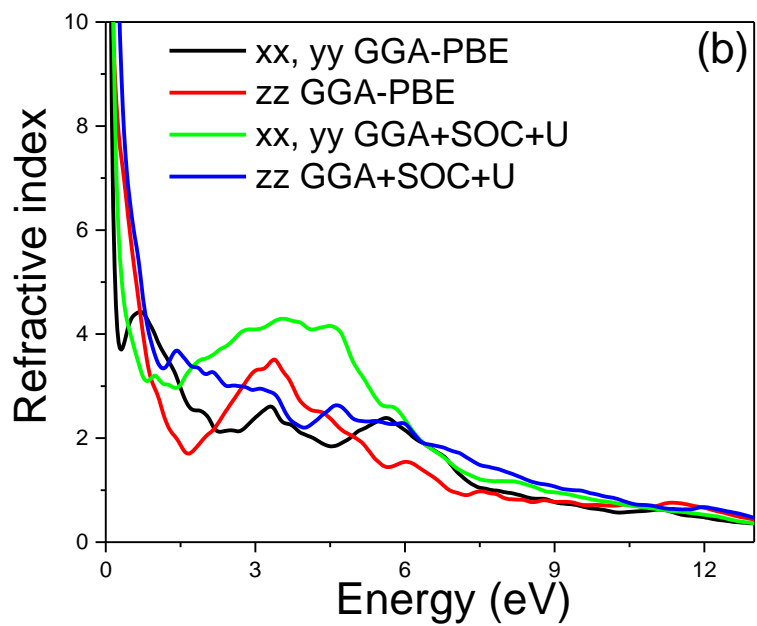

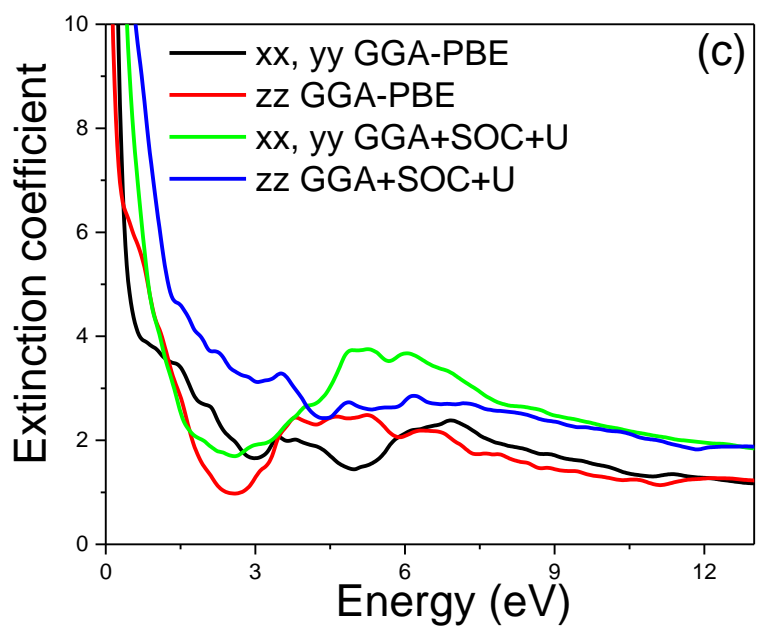

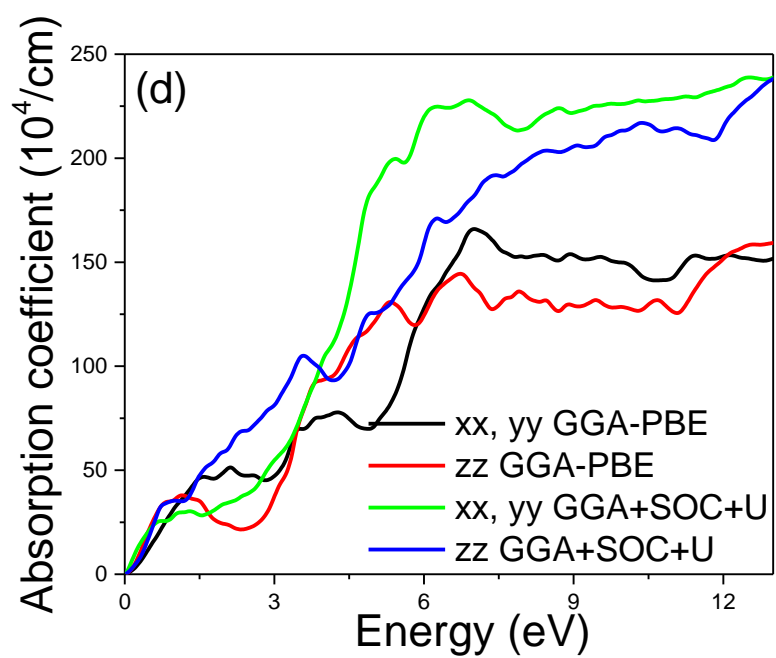

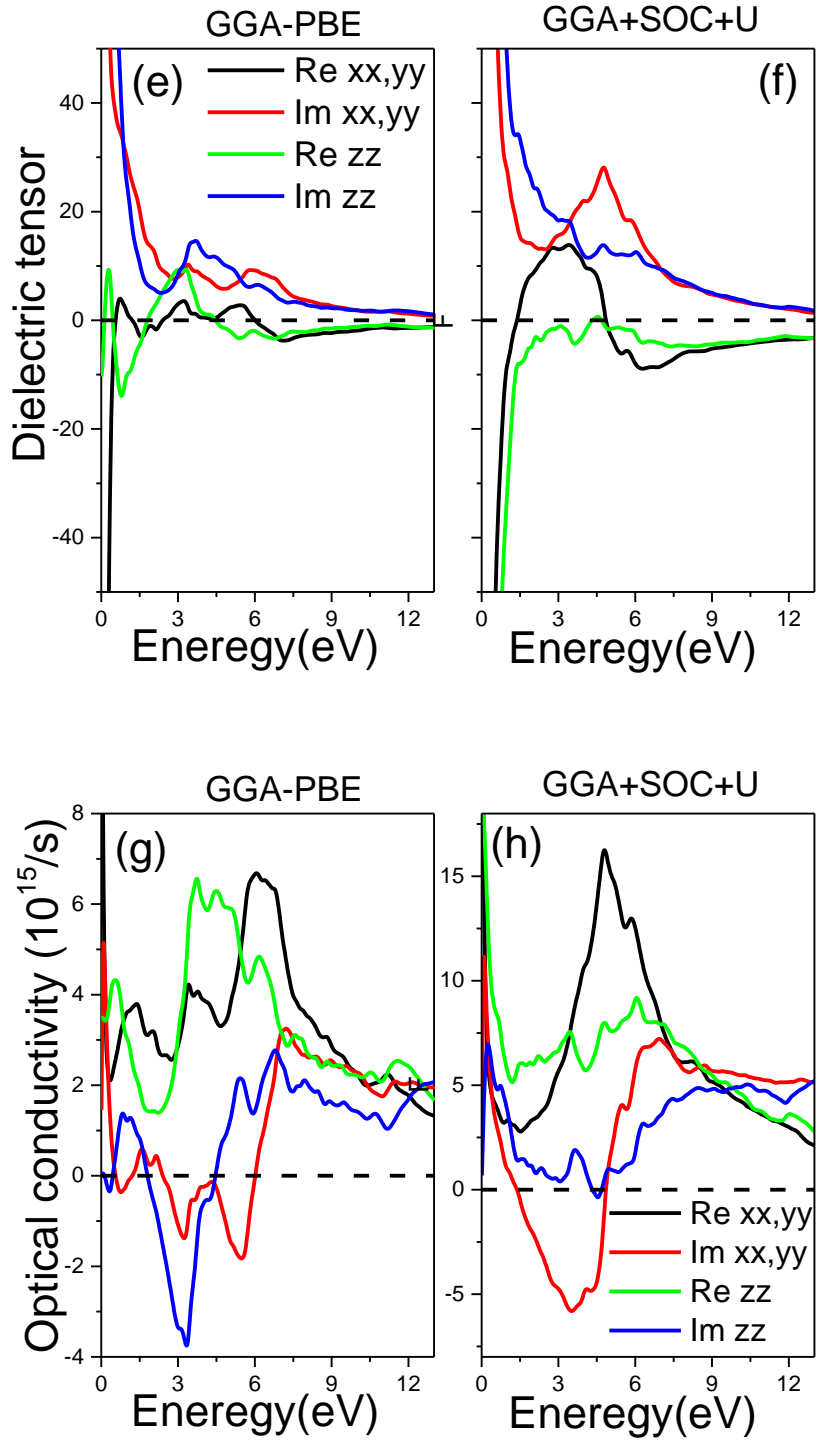

**Fig.5**: *Optical properties of EuCo$_2$As$_2$: the electron energy loss (a), the refractive index (b), the extinction coefficient (c), the absorption coefficient (d), the dielectric tensor (e), and the optical conductivity (f) by using the GGA-PBE and GGA+SOC+U approaches.*

### 3.6. Thermodynamic properties

Thermodynamic characteristics play a crucial role in the field of material science and engineering. In this particular study, the thermal properties of EuCo$_2$As$_2$ are investigated using the quasi-harmonic Debye model, considering temperature ranges from 0 to 1800 K and strain ranges from 0 to 12 GPa. The Debye temperature (θ) is a key parameter in this model, as it provides insights into various physical properties of solids, including heat capacity, elastic factors, thermal expansion rate, and entropy, and is closely related to specific heat, structural stability, lattice vibrations, and chemical bond strength [30]. The Debye temperature (θ) has been computed, from the elastic constants, using the relation:

$$\theta = \frac{\hbar}{k_B}\left(\frac{6\pi^2}{V_{at}}\right)^{1/3}\left[\frac{1}{3}\left(\frac{3K+4G}{3\rho}\right)^{-3/2} + \frac{2}{3}\left(\frac{G}{\rho}\right)^{-3/2}\right] \quad (2)$$

$V_{at}$ is the volume per atom, $\rho$ is the density, and K and G are the bulk and shear modulus, respectively. The Debye temperature was calculated from the full elastic constant tensor averaged by the Reuss–Voigt–Hill scheme (K = 192.6 GPa, G = 99.2 GPa) using equation (2). Fig. 6(a) illustrates the impact of temperature and pressure on the volume of EuCo$_2$As$_2$. It can be observed that as the temperature increases, the volume also increases, reaching a higher value of 91.3 Å$^3$ at 1000 K. Conversely, the application of pressure leads to a decrease in volume. In Fig. 6(b), the relationship between the Debye temperature and tensile strain has been depicted. It is evident that the Debye temperature increases proportionally with increasing tensile strain. Additionally, the curves of the Debye temperature exhibit a decreasing trend with increasing temperature. These findings indicate that the introduction of dilatation strain can enhance the mechanical and thermal stability of EuCo$_2$As$_2$. Within the Debye model, the Grüneisen parameter could be calculated using the quasi-harmonic relation:

$$\gamma = -\frac{d\ln\theta}{d\ln V} \quad (3)$$

The Grüneisen parameter under pressure for EuCo$_2$As$_2$ is presented in Fig. 6(c). At lower pressures (0-2 GPa), the Grüneisen parameter increases in proportion to the temperature. However, when higher pressures (4-12 GPa) are applied, the Grüneisen parameter exhibits a decrease with temperature, starting from higher values at lower temperatures. Overall, the utilization of the quasi-harmonic Debye model provides valuable insights into the thermodynamic behavior of EuCo$_2$As$_2$, enabling a deeper understanding of its thermal properties, structural stability, and response to external factors such as temperature and pressure.

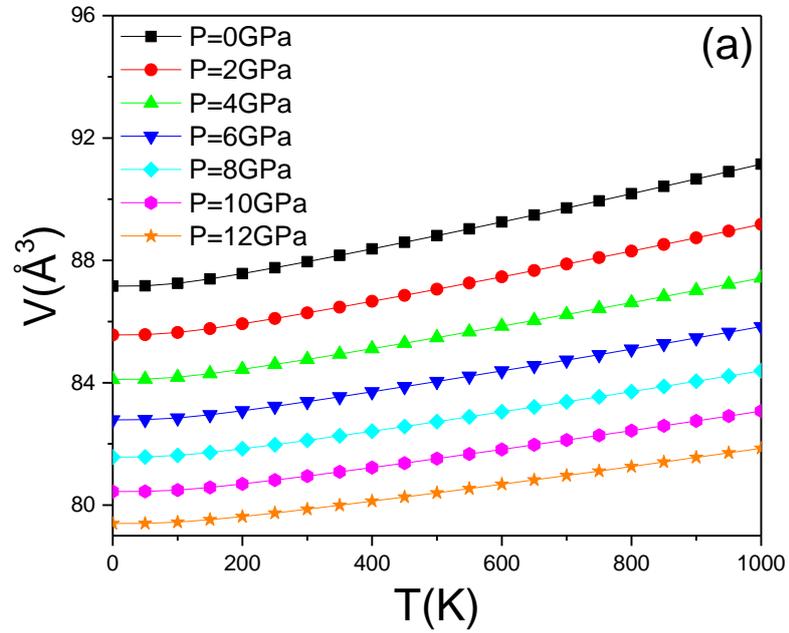

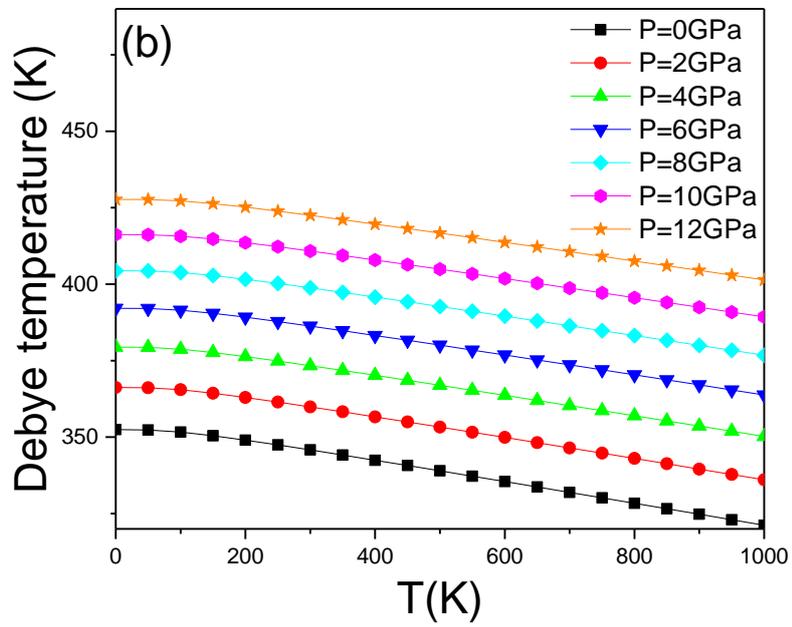

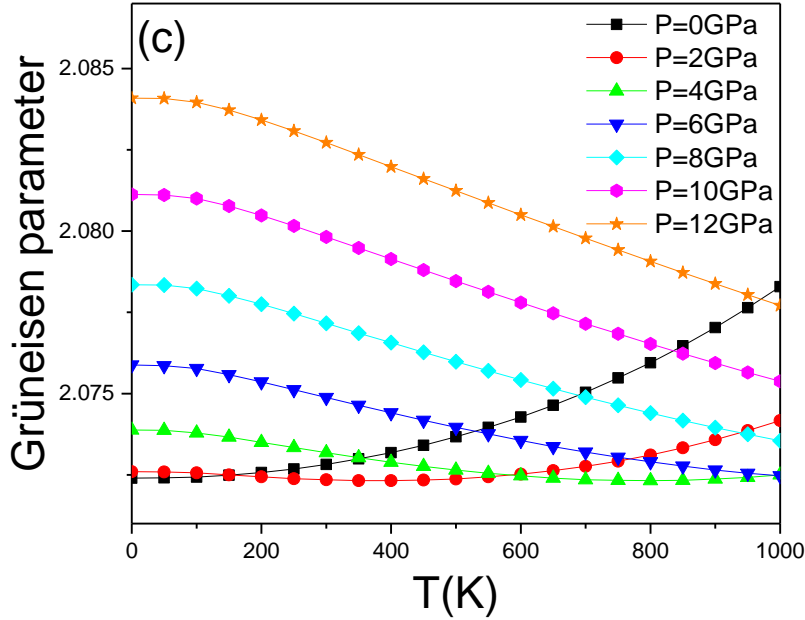

*Fig. 6: The volume (a), the Debye temperature (b) and the Grüneisen parameter (c) of EuCo$_2$As$_2$ under different strains.*

### 3.7. Thermoelectric properties

The thermoelectric properties of materials are known to be highly influenced by the details of their band structure. In this study, we have employed a combination of electronic structure calculations and Boltzmann transport theory [31], implemented in the BoltzTraP code, to analyze and predict the thermoelectric performance of EuCo$_2$As$_2$. The BoltzTraP code utilizes an interpolated band structure to compute the required derivatives for evaluating thermoelectric properties under the constant relaxation time approximation. It is important to note that the temperature dependence of the energy band structure is neglected in this analysis. The thermoelectric performance of EuCo$_2$As$_2$ is characterized by several key parameters, including the Seebeck coefficient (S), electrical conductivity ($\sigma$), electronic thermal conductivity ($\kappa_e$), and lattice thermal conductivity ($\kappa_L$). We calculate the transport properties of EuCo$_2$As$_2$ for both spin-up and spin-down configurations using the GGA-PBE and GGA+SOC+U approaches across a temperature range from 100 K to 1800 K (refer to Fig. 7). Fig. 7(a) illustrates the variation of the Seebeck coefficient (S) with temperature for EuCo$_2$As$_2$. Our observations indicate that the Seebeck coefficient decreases as the temperature increases, regardless of the approach used (GGA-PBE and GGA+SOC+U) and the spin channel. The results suggest that the material exhibits an n-type behavior, as evidenced by the negative values of the Seebeck coefficient. To gain a comprehensive understanding of the thermoelectric behavior of the alloy for practical applications, it is essential to consider the overall or total Seebeck coefficient. By integrating electronic structure calculations and Boltzmann transport theory, we are able to analyze and predict the thermoelectric performance of EuCo$_2$As$_2$, shedding light on its suitability for thermoelectric applications. Applying the two-current method [32], we have estimated the total Seebeck coefficient as follows:

$$S = \frac{S(\uparrow)\sigma(\uparrow) + S(\downarrow)\sigma(\downarrow)}{\sigma(\uparrow) + \sigma(\downarrow)} \qquad (4)$$

Fig. 7(b) shows the electrical conductivity by relaxation time ($\sigma/\tau$) versus temperature for EuCo$_2$As$_2$ for both spin states using GGA-PBE and GGA+SOC+U approaches. It can be seen that the electrical conductivity by relaxation time ($\sigma/\tau$) increases with temperature for the spin-up and spin-down, respectively. Therefore, the electric conductivity of the n-type doping system is greater than that of the p-type in this case at the given temperature. Moreover, we can observe that at 1800 K, the value of the electric conductivity by relaxation time using the GGA+SOC+U approach is $5.25 \times 10^{20}$ ($\Omega^{-1}.m^{-1}.s^{-1}$). The present results demonstrate strong electric conductivity and weak resistivity. Consequently, we have electrical charge transport with very small losses caused by the Joule effect, which represents a significant benefit for it to be a suitable thermoelectric substance.

$$\sigma = \sigma(\uparrow) + \sigma(\downarrow) \qquad (5)$$

The thermal conductivity κ is composed of the electronic part $\kappa_e$ and the lattice (phonon) part $\kappa_L$. Firstly, we calculate the electronic by relaxation time part ($\kappa_e/\tau$) of the thermal conductivity of EuCo$_2$As$_2$ for both spin up and down versus temperature by utilizing the BoltzTraP package [23]. In Fig. 7(c), we present the response of the electronic thermal conductivity ($\kappa_e/\tau$) by relaxation time ($\tau$) versus temperature between 0 K and 1800 K for EuCo$_2$As$_2$. We can clearly observe the electronic thermal conductivity increasing with the temperature to reach $20 \times 10^{15}$ (W.m$^{-1}$.K$^{-1}$.s$^{-1}$) and $17.33 \times 10^{15}$ (W.m$^{-1}$.K$^{-1}$.s$^{-1}$) for the GGA+SOC+U and GGA-PBE approaches, respectively, at 1800 K.

The lattice thermal conductivity $\kappa_L$ is obtained by using the temperature of Debye (θ), and the Grüneisen coefficient (γ) are obtained from computations via the Gibbs2 software programs [33]. The methodology also adopts the implementation of an exceptional relationship to calculate the part of the thermal conductivity $\kappa_L$ by using the Slack approximation outlined by the equation

$$\kappa_L = \frac{A\theta_D^3 V^{1/3} M}{\gamma^2 n^{2/3} T} \quad (6)$$

Where A is a group of physical parameters. According to [34, 35, 36], this can be computed as

$$A = \frac{2.43 \times 10^{-8}}{1 - 0.514/\gamma + 0.228/\gamma^2} \quad (7)$$

$\theta_D$ is the temperature of Debye, $\gamma$ is the Grüneisen coefficient, V is the volume by atom, T is the temperature, n is the number of atoms in the primitive unit cell, and M is the atomic mass. The thermal conductivity of the lattice ($\kappa_L$) for the EuCo$_2$As$_2$ using GGA-PBE approaches for both channels, in Fig. 7(d), decreases exponentially with an increase in temperature and with an increase in pressure. The high value of $\kappa_L$ identifies the presence of an anti-harmonic effect in the material. Regarding the small lattice thermal conductivity value, we may state that the current material may be a very promising material for thermoelectric utility.

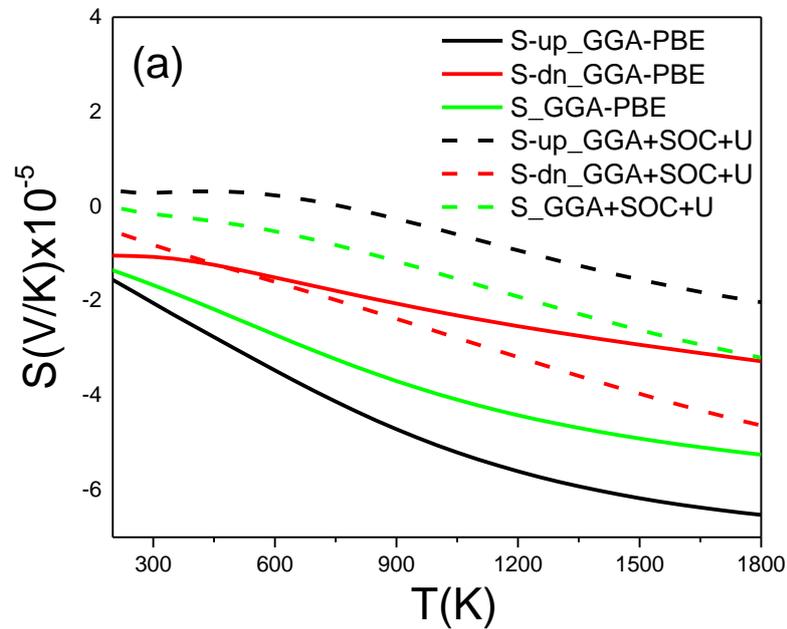

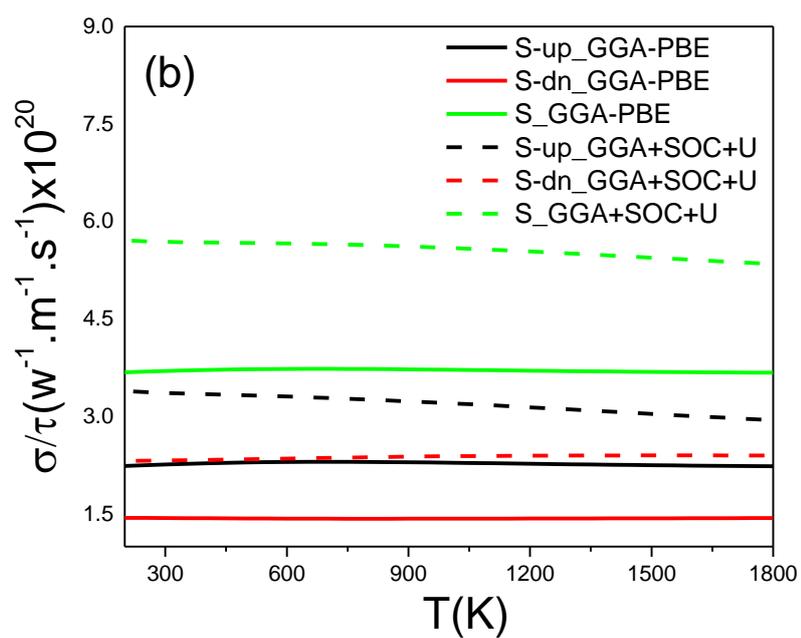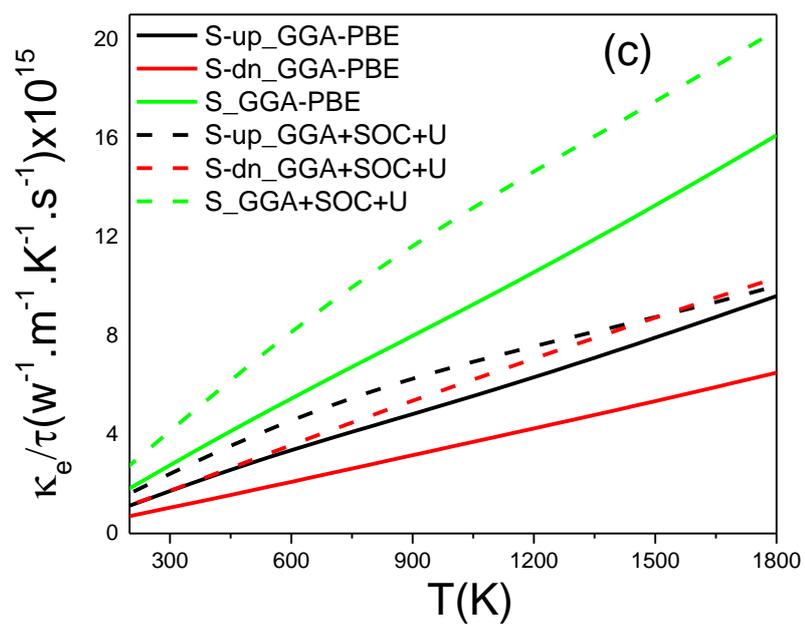

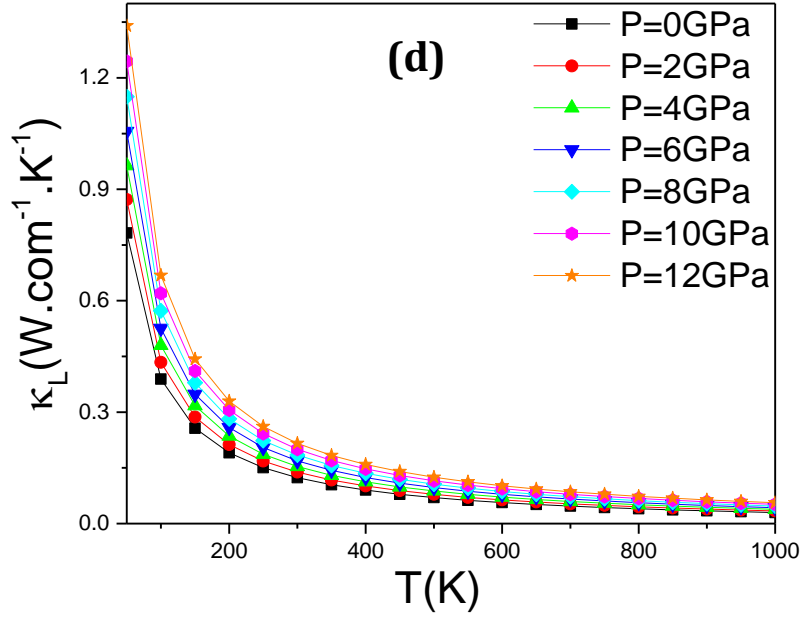

*Fig. 7: The Seebeck coefficient (a), the electrical conductivity (b), the electronic part of the thermal conductivity (c), and the thermal conductivity of the lattice (d) of EuCo$_2$As$_2$ under different strains using GGA-PBE and GGA+SOC+U approaches.*

## 4. Conclusion

In conclusion, the structural, elastic, optical, thermoelectric and thermodynamic properties of EuCo$_2$As$_2$ have been studied using ab initio computations. Based on the electronic properties, the metallic character of the compound is demonstrated and the ferromagnetic phase is shown to be energetically more favorable. For the first time, we have reported elastic constants, the Grüneisen parameter and the Debye temperature of this compound. The results of the optical analysis indicate that EuCo$_2$As$_2$ has adequate qualities as light absorber, with a reasonable optical band gap and a high capacity to absorb light. This is supported by absorption, optical conductivity, and dielectric tensor calculations. The thermoelectric properties conclude that the material exhibits an n-type behavior, as evidenced by the negative values of the Seebeck coefficient. Regarding the small lattice thermal conductivity value of $\kappa_L$ = *0.10 (W.cm$^{-1}$.K$^{-1}$)*, we may state that the current material may be a very promising material for thermoelectric utility. To the best of our knowledge, the majority of the researched parameters have not yet been published by other researchers. We included them in the goal of inspiring more research.